\documentclass[12pt]{article}
\usepackage{amsmath}
\usepackage{breqn}
\usepackage{cite}

\begin{document}

\title{\bf Modified Friedmann equations from fractional entropy}

\author{{Zeynep \c{C}oker$^{a}$ \thanks{Email: cokerzeynep@gmail.com}\hspace{1mm},    \"{O}zg\"{u}r \"{O}kc\"{u}$^{a,b}$ \thanks{Email: ozgur.okcu@ogr.iu.edu.tr}\hspace{1mm},
Ekrem Aydiner$^{a,b}$ \thanks{Email: ekrem.aydiner@istanbul.edu.tr}} \\
$^a${\small {\em Department of
	Physics, Faculty of Science, Istanbul University,
		}}\\{\small {\em Istanbul, 34134, Türkiye}} \\
  $^b${\small {\em Theoretical and Computational Physics Research Laboratory, Istanbul University,
		}}\\{\small {\em Istanbul, 34134, Türkiye}} }
\maketitle

\begin{abstract}
Based on the fractional black hole entropy (Jalalzadeh S. et al., Eur. Phys. J. C, 81 (2021) 632), we derive the modified Friedmann equations from two different frameworks. First, we consider the modifications of Friedmann equations from the first law of thermodynamics at the apparent horizon. We show that the generalized second law (GSL) of thermodynamics always holds in a region bounded by the apparent horizon. Then, we obtain Friedmann equations from Verlinde's entropic gravity framework. We also compute the fractional corrections to the deceleration parameter $q$ in the flat case $k=0$ for both frameworks. Furthermore, we consider the time to reach the initial singularity for the two frameworks. The results indicate that  the initial singularity is accessible for both frameworks. However, fractional effects may provide a constraint on the equation of state parameter in the entropic gravity scenario since the time is imaginary for $-2/3\alpha<\omega<-1/3$. 
\end{abstract}

\section{Introduction}
Black holes thermodynamics is one of the most promising research fields in theoretical physics since it reveals the multicultural character of gravity, namely, the deep connection between gravitation, quantum mechanics, and thermodynamics \cite{Bekenstein1972,Bekenstein1973,Bardeen1973,Hawking1974,Bekenstein1974,Hawking1975}. Black hole surface area $A$ and surface gravity $\kappa$ correspond to thermodynamic quantities entropy $S$ and temperature $T$, respectively.  Inspired by black hole thermodynamics, Jacobson derived the Einstein field equations from the first law of thermodynamics \cite{Jacobson1995}. He obtained the field equations by  considering the entropy-area expression with Clausius relation, $\delta Q=TdS$, under the assumption of the relation is valid for local Rindler causal horizons through each spacetime point. Here, $\delta Q$ and $T$ correspond to energy flux and Unruh temperature, respectively.  Following Jacobson's seminal work, there have been many papers aimed to  reveal the deep connection between gravitational dynamics and horizon thermodynamics \cite{Padmanabhan2002,Eling2006,Paranjape2006,Kothawala2007,Padmanabhan2007,Cai2005,Akbar2006,Akbar2007,Cai2007a,Cai2007b,Cai2008,Sheykhi2010a,Awad2014,Salah2017,Okcu2020,Sheykhi2010b,Sheykhi2007a,Sheykhi2007b,Sheykhi2009,Sheykhi2018,Nojiri2019,Lymperis2018,Karami2011,Sheykhi2019,Saridakis2020,Barrow2021,Saridakis2021,Anagnostopoulos2020,Sheykhi2021,Asghari2021,Sheykhi2022,Sheykhi2022b,Asghari2022,Sheykhi2023,Sheykhi2023b,Lymperis2021,Drepanou2022,Odinstov2023,Abreu2022,Verlinde2011,Cai2010,Shu2010,Sheykhi2010c,Sheykhi2011,Gao2010,Ling2010,FCai2010,Hendi2011,Basilakos2012,Senay2021,Sefiedgar2017,Feng2018,Abreu2018,Jusufi2023,Jusufi2023b}.  For example, the studies handling the relation between Einstein field equations and the first law of thermodynamics can be found in refs.~\cite{Padmanabhan2002,Eling2006,Paranjape2006,Kothawala2007,Padmanabhan2007}.
Motivated by Jacobson's study, Cai and Kim \cite{Cai2005} derived higher-dimensional Friedmann equations from the first law in the form of $-dE=T_{h}dS_{h}$ at 
apparent horizon. Here $-dE$ corresponds to the energy flux passing through the apparent horizon for the infinitesimal time interval at a fixed horizon radius. The temperature and entropy at the apparent horizon are given by \cite{Cai2005}
\begin{equation}
\label{Temp-entropy1}
T_{h}=\frac{1}{2\pi\tilde{r_{A}}},\qquad\qquad S_{h}=\frac{A}{4},
\end{equation}
where $A$ and $\tilde{r_{A}}$ are area and apparent horizon, respectively \footnote{We use the units $\hbar=c=G_{N}=L^2_{Pl}=1$ throughout the paper.}. Moreover, they also obtained Friedmann equations from the entropy formulae of Gauss-Bonnet gravity and Lovelock gravity theories, where the entropy is not proportional to the horizon area. Based on ref.~\cite{Cai2005},  Friedmann equations in the scalar$-$tensor and $f(R)$ gravity theories were derived in ref.~\cite{Akbar2006}. Although one can obtain the Friedmann equations from the above equations, this framework has some shortcomings such as the limitation on the equation$-$of$-$state only vacuum$-$energy$-$dominated or de Sitter spacetime. Besides, the horizon temperature is not proportional to the surface gravity $\kappa$. Assuming the proportionality of temperature and surface gravity in addition to the entropy-area relation, one should update the first law of thermodynamics at the apparent horizon as follows \cite{Akbar2007}:
\begin{equation}
\label{firstLaw}
dE=T_{h}dS_{h}+WdV,
\end{equation}
where $W$ is the work density and $E=\rho V$ is the total energy in volume $V$ enclosed by the apparent horizon. Then, following ref.~\cite{Akbar2007}, thermodynamics of the apparent horizon and unified first law were also studied in refs.~\cite{Cai2007a,Cai2007b}. Recently, Friedmann equations and apparent horizon thermodynamics have been widely studied in the literature ~\cite{Cai2008,Sheykhi2010a,Awad2014,Salah2017,Okcu2020,Sheykhi2010b,Sheykhi2007a,Sheykhi2007b,Sheykhi2009,Sheykhi2018,Nojiri2019,Lymperis2018,Karami2011,Sheykhi2019,Saridakis2020,Barrow2021,Saridakis2021,Anagnostopoulos2020,Sheykhi2021,Asghari2021,Sheykhi2022,Sheykhi2022b,Asghari2022,Sheykhi2023,Sheykhi2023b,Lymperis2021,Drepanou2022,Odinstov2023,Abreu2022}.

Another interesting aspect of thermodynamical gravitation is Verlinde's entropic gravity \cite{Verlinde2011}. In 2010, Verlinde proposed that gravity is not a fundamental force since combining gravity with quantum mechanics is harder than other forces. He claimed that gravity is interpreted as an entropic force that emerged due to the entropy changes of bits on the holographic screen. Assuming a holographic screen with an Unruh temperature, he derived Newton's second law. Furthermore, using the holographic principle and the equipartition law of energy, he also derived Newton's gravitational law and Einstein field equations. Subsequently, many studies on the derivations of Newton's gravitational law, Einstein field equations and Friedmann equations in entropic gravity have been published \cite{Cai2010,Shu2010,Sheykhi2010c,Sheykhi2011,Gao2010,Ling2010,FCai2010,Hendi2011,Basilakos2012,Senay2021,Sefiedgar2017,Feng2018,Abreu2018,Jusufi2023,Jusufi2023b}.

It is widely known that entropy can be modified in the context of various theories. The literature is rich with the applications of modified Friedmann equations for both the first law at the apparent horizon and entropic gravity. Motivated by various  approaches, loop quantum gravity \cite{Cai2008,Sheykhi2010a}, generalised uncertainty principle \cite{Awad2014,Salah2017,Okcu2020}, rainbow gravity \cite{Sefiedgar2017,Feng2018} corrected Friedmann equations were derived. Moreover, Tsallis ~\cite{Sheykhi2018,Nojiri2019,Lymperis2018}, Barrow ~\cite{Saridakis2020,Barrow2021,Anagnostopoulos2020,Saridakis2021,Sheykhi2021,Asghari2021,Sheykhi2022,Asghari2022,Sheykhi2023,Sheykhi2023b} and Kaniadakis ~\cite{Lymperis2021,Drepanou2022,Abreu2018}  entropies modified Friedmann equations can be found in the literature.

Recently, Jalalzadeh et al. investigated the effects of fractional quantum mechanics (FQM) on Schwazrschild black hole thermodynamics \cite{Jalalzadeh2021}. By using a space-fractional derivative of second order (Riesz derivative), they obtained fractional black hole entropy from a modified Wheeler-DeWitt equation. In order to obtain the modified Wheeler-DeWitt equation, they first considered the canonical quantization procedure of the Schawarzschild black hole and obtained the corresponding Hamiltonian. Then, including the quantum Riesz derivative in the momentum operator of the Hamiltonian leads to the fractional Wheeler-deWitt equation. The fractional entropy is given by\cite{Jalalzadeh2021}
  \begin{equation}
 \label{entropy}
S_{h}=\left(\pi\tilde{r_{A}}^{2}\right)^{\frac{2+\alpha}{2\alpha}}, \qquad 1<\alpha\leq2,
 \end{equation}
where $\alpha$ is the fractional parameter and the standard case is recovered for $\alpha=2$. This equation implies that the entropy is a power$-$law function of its area. We note that this entropy resembles the Barrow \cite{Barrow2020}  and Tsallis \cite{Tsallis2013} entropies although they have different motivations and physical principles. 

In this work, we would like to investigate the modifications of the Friedmann equations for fractional entropy. In order to obtain the Friedmann equations, we consider two frameworks, namely, the first law of thermodynamics at the apparent horizon and entropic gravity. The fractional derivative extends the order of derivative reel or complex numbers. There are various kinds of fractional derivatives, Liouville, Riemann, Caputo, Riesz fractional derivatives, etc. \cite{Herrmann2011}. None of these derivatives successfully explain the all experimental data. Instead, their agreements with the experiments depend on the specific problems. Fractional calculus especially finds many places in the applications of quantum mechanics. The fractional generalization of quantum mechanics is known as FQM \cite{Laskin2000,Laskin2002,Laskin2000b,Laskin2000c,Laskin2017,Wang2007,Laskin2018,Laskin2010}. The interested reader may refer to Laskin's monograph on FQM in ref.~\cite{Laskin2018} and review in ref.~\cite{Laskin2010}. Moreover, recently many studies devoted to relativistic gravitation and cosmology have been considered in FQM \cite{Moniz2020,Moniz2020b,Calcagni2017,Calcagni2018,Calcagni2010,Calcagni2012,Calcagni2013,Calcagni2012b,Calcagni2017b,Amelino2017,Torres2020,Roberts2009,Nabulsi2017,Landim2021,Jamil2012,Shchigolev2011,Shchigolev2021,Nabulsi2021,Nabulsi2018,Jalalzadeh2022,Garcia2022,Rasouli2022,González2023,Riascos2023,Socorro2023,Jawada2019,Asghari2022b,Asghari2022c,Aydiner2022}.

The paper is organized as follows: In the next section, we obtain the fractional  Friedmann equations from the first law of thermodynamics at the apparent horizon. We investigate the deceleration parameter and time to reach the initial singularity. Then, we check the validity of GSL. In the third section, we derived fractional Friedmann equations from the entropic gravity framework. Similarly, we study the deceleration parameter and calculate the time to reach the singularity. Finally, the conclusions are presented in the last section.

\section{Friedmann equations from the first law of thermodynamics}

Let us begin to take a quick glimpse at the basic elements of the Friedmann-Robertson-Walker (FRW) universe. The line element of the FRW universe in the compact form is defined by \cite{Cai2005}
 \begin{equation}
 \label{lineElement}
 ds^{2}=h_{ab}dx^{a}dx^{b}+\tilde{r}d\Omega^{2},
 \end{equation}
where $\tilde{r}=a(t)r$, $a(t)$ is the scale factor, $x^{a}=(t,r)$, and $h_{ab}=diag\left(-1,a^{2}/(1-kr^{2})\right)$ is the two$-$dimensional metric. $k=$ $-1$, $0$, and $1$ correspond to the open, flat, and closed universe, respectively. The apparent horizon is defined by \cite{Cai2005}
 \begin{equation}
 \label{apparentHor}
 \tilde{r_{A}}=ar=\frac{1}{\sqrt{H^{2}+k/a^{2}}},
 \end{equation}
 where $H=\dot{a}/a$ is the Hubble parameter, and dot denotes the derivative with respect to time. The surface gravity of the horizon is given by \cite{Cai2005,Hayward1998}
\begin{equation}
\label{kappa}
\kappa=-\frac{1}{\tilde{r_{A}}}\left(1-\frac{\dot{\tilde{r_{A}}}}{2H\tilde{r_{A}}}\right),
\end{equation}
and the corresponding temperature of the apparent horizon is given by \cite{Akbar2007}
\begin{equation}
\label{Temp}
T_{h}=\frac{\kappa}{2\pi}=-\frac{1}{2\pi\tilde{r_{A}}}\left(1-\frac{\dot{\tilde{r_{A}}}}{2H\tilde{r_{A}}}\right).
\end{equation}
We assume the matter and energy of the universe as an ideal fluid, thus the corresponding energy-momentum tensor is given by
  \begin{equation}
 \label{energyMomentumTensor}
 T_{\mu\nu}=(\rho+p)u_{\mu}u_{\nu}+pg_{\mu\nu},
 \end{equation}
where $\rho$, $p$ and $u^{\mu}$ are energy density, pressure, and four$-$velocity of the fluid, respectively. From the conservation of energy-momentum tensor, i.e., $T_{;\beta}^{\alpha\beta}=0$, one obtains the continuity equation as 
 \begin{equation}
 \label{continuityEqu}
 \dot{\rho}+3H(\rho+p)=0.
 \end{equation}
According to the arguments of ref.~\cite{Hayward1998}, the work density is defined by
 \begin{equation}
 \label{workDensity}
 W=-\frac{1}{2}T^{ab}h_{ab}=\frac{1}{2}(\rho-p).
 \end{equation}
At this point, we briefly mention eq.~(\ref{workDensity}). Here, $W$ is the work done by the volume change of the universe. 

Now, we start to calculate eq.~(\ref{firstLaw}) step by step. Employing the volume \cite{Akbar2007}
\begin{equation}
 V=\frac{4}{3}\pi \tilde{r_{A}}^{3},
\label{volume}
\end{equation}
the total energy of universe $E=\rho V$ and eq.~(\ref{continuityEqu}), one finds the differential of $E$ as follows
 \begin{equation}
 \label{dE}
 dE=\rho dV+Vd\rho=4\pi\rho\tilde{r_{A}}^{2}d\tilde{r_{A}}-4\pi(\rho+p)\tilde{r_{A}}^{3}Hdt,
 \end{equation}
and from eqs.~(\ref{workDensity}) and (\ref{volume}), $WdV$ is given by
\begin{equation}
\label{WdV}
WdV=2\pi(\rho-p)\tilde{r_{A}}^{2}d\tilde{r_{A}}.
\end{equation}
Differentiating the entropy in eq.~(\ref{entropy}), we obtain $T_{h}dS_{h}$ as
 \begin{equation}
 \label{TdS}
 T_{h}dS_{h}=-\left(\frac{2+\alpha}{2\alpha}\right)\left(1-\frac{\dot{\tilde{r_{A}}}}{2H\tilde{r_{A}}}\right)\pi^{\frac{2-\alpha}{2\alpha}}\tilde{r_{A}}^{\frac{2-\alpha}{\alpha}}d\tilde{r_{A}}.
 \end{equation}
Amalgamating eqs.~(\ref{dE}), (\ref{WdV}) and (\ref{TdS}) in eq.~(\ref{firstLaw}) and employing the relation
\begin{equation}
\label{difAppa}
d\tilde{r_{A}}=-H\tilde{r_{A}}^{3}\left(\dot{H}-\frac{k}{a^{2}}\right)dt,
\end{equation}
we obtain
\begin{equation}
 \label{dynamicalEquation}
4\pi(\rho+p)Hdt=\left(\frac{2+\alpha}{2\alpha}\right)\pi^{\frac{2-\alpha}{2\alpha}}\tilde{r_{A}}^{\frac{2-4\alpha}{\alpha}}d\tilde{r_{A}}.
 \end{equation}
 Combining the continuity equation with the above equation and integrating the above equation gives the first Friedmann equation. Then using  eq.~(\ref{difAppa}) in the above equation yields the the second Friedmann equation. The modified Friedmann equations are given by
  \begin{eqnarray}
 \label{Fried1}
 \left(\frac{2+\alpha}{2\alpha}\right)\pi^{\frac{2-\alpha}{2\alpha}}\tilde{r_{A}}^{\frac{2-3\alpha}{\alpha}}=\frac{4\pi}{3}\left(\frac{3\alpha-2}{\alpha}\right)\rho
 \nonumber, \\
 \left(\frac{2+\alpha}{2\alpha}\right)\pi^{\frac{2-\alpha}{2\alpha}}\tilde{r_{A}}^{\frac{2-\alpha}{\alpha}}\left(\dot{H}-\frac{k}{a^{2}}\right)=-4\pi(\rho+p),
 \end{eqnarray}
where we set the integration constant to zero in the first equation \footnote{ Alternatively, deriving the first Friedmann equation and employing the continuity equation gives the second Friedmann equation.}. These modified equations are reduced usual form in the limit $\alpha\rightarrow2$. Using eq.~(\ref{apparentHor}), Friedmann equations are expressed in terms of Hubble parameter, i.e.,
 \begin{eqnarray}
 \label{Fried2}
\left(H^{2}+\frac{k}{a^{2}}\right)^{\frac{3\alpha-2}{2\alpha}}=\frac{8\pi}{3}\left(\frac{3\alpha-2}{(2+\alpha)\pi^{\frac{2-\alpha}{2\alpha}}}\right)\rho,\nonumber \\
\left(\frac{2+\alpha}{2\alpha}\right)\pi^{\frac{2-\alpha}{2\alpha}}\left(H^{2}+\frac{k}{a^{2}}\right)^{\frac{\alpha-2}{2\alpha}}\left(\dot{H}-\frac{k}{a^{2}}\right)=-4\pi(\rho+p).
 \end{eqnarray}

 Now, we would like to investigate the fractional effects on the deceleration parameter $q$ for $k=0$. It is given by
 \begin{equation}
q=-\frac{a\ddot{a}}{\dot{a}^{2}},
\label{decelPar}
 \end{equation}
where positivity and negativity of $q$ mean decelerated and accelerated phases, respectively. From the continuity  equation (\ref{continuityEqu}), we obtain 
\begin{equation}
\rho=\rho_{0}a^{-3(1+\omega)},
\label{RhoScaleRel}
\end{equation}
where we use $p=\omega\rho$ as equation of state and $\rho_{0}$ is a constant. Substituting eq.~(\ref{RhoScaleRel}) in the first equation in (\ref{Fried2}) yields the following solution
\begin{equation}
a(t)\propto t^{\frac{3\alpha-2}{3\alpha(\omega+1)}}.
\label{solFora}
\end{equation}
From the above solution, the deceleration parameter is given by
\begin{equation}
q=\frac{2+3\alpha\omega}{3\alpha-2}.
\label{decelParC1}
\end{equation}
For radiation $(\omega=1/3)$ and matter $(\omega=0)$ dominated eras, deceleration parameters are given by
\begin{equation}
q_{rad}=\frac{2+\alpha}{3\alpha-2},\qquad q_{m}=\frac{2}{3\alpha-2},
\label{decelParMat1}
\end{equation}
respectively. We recover the standard forms of eqs.~(\ref{decelParMat1}) in the limit $\alpha\rightarrow2$. Since $1<\alpha\leq2$, both deceleration parameters are always positive, i.e., matter$-$ and radiation-dominated eras correspond to decelerated phases. The results imply that fractional effects do not provide an alternative to dark energy. This is in contrast to Tsallis cosmology \cite{Sheykhi2018} while it is similar to Barrow cosmology \cite{Sheykhi2022}. At the late time acceleration, $q$ is negative for $\omega<-\frac{2}{3\alpha}$. It clearly shows a shift from the standard case, i.e., $\omega<-1/3$.

\begin{figure}
\centerline{\includegraphics[width=11cm]{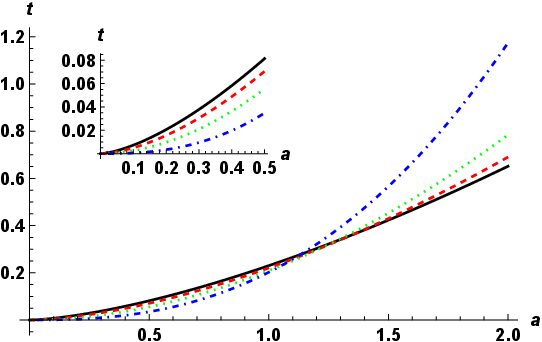}}
\caption{Time to reach  singularity vs scale factor. Dashed$-$dotted blue line $\alpha=1.1$, dotted green line $\alpha=1.4$, dashed red line $\alpha=1.7$ and black solid line $\alpha=2$. We take $\rho_{0}=1$ and $\omega=0$.}
\label{sing1}
\end{figure}

Now, we prospect whether the initial singularity of the universe is accessible or not. For this purpose, we use the analysis in refs.~\cite{Awad2013,Hendi2017}. Combining the continuity equation with the first Friedmann equation for $k=0$, we get
\begin{equation}
\dot{\rho}=\pm3(\omega+1)\sqrt{\pi}\left(\frac{8(3\alpha-2)}{3(\alpha+2)}\right)^{\frac{\alpha}{3\alpha-2}}\rho^{\frac{4\alpha-2}{3\alpha-2}}.
\label{singularityProspect1a}
\end{equation}
Integrating this equation from a finite density $\rho^{*}$ to an infinite one, we find
\begin{equation}
t=\pm\frac{1}{3\sqrt{\pi}(\omega+1)}\left(\frac{8(3\alpha-2)}{3(\alpha+2)}\right)^{\frac{\alpha}{2-3\alpha}}\int_{\rho^{*}}^{\infty}\rho^{\frac{4\alpha-2}{2-3\alpha}}d\rho,
\label{singularityProspect1b}
\end{equation}
and the solution is given by
\begin{equation}
t=\pm\frac{1}{3\sqrt{\pi}(\omega+1)}\left(\frac{8(3\alpha-2)}{3(\alpha+2)}\right)^{\frac{\alpha}{2-3\alpha}}\left(\frac{2-3\alpha}{\alpha}\right)\rho_{*}^{\frac{\alpha}{2-3\alpha}}.
\label{singularityProspect1c}
\end{equation}
The result implies that the Big Bang singularity is accessible since the time to reach singularity is finite. Moreover, we can make the analysis more specific by substituting eq.~(\ref{RhoScaleRel}) into the integral in eq.~(\ref{singularityProspect1b}). We obtain
\begin{equation}
t=\frac{\pm1}{3(1+\omega)\sqrt{\pi}}\left(\frac{8\rho_{0}(3\alpha-2)}{3(\alpha+2)}\right)^{\frac{\alpha}{2-3\alpha}}\left(\frac{3\alpha-2}{\alpha}\right)a^{\frac{3\alpha(1+\omega)}{3\alpha-2}}\Biggr|_{a_{*}}^{0},
\label{singularityProspect1d}
\end{equation}
where $a_{*}$ is a finite and nonzero scale factor. It is straightforward to find $t=\infty$ for the condition $\omega<-1$. However, this condition is independent of the fractional case since the standard case ($\alpha=2$) similarly,
\begin{equation}
t=\frac{\pm2}{3(1+\omega)}\left[\frac{8\pi\rho_{0}}{3}\right]^{-\frac{1}{2}}a^{\frac{3(1+\omega)}{2}}\Biggr|_{a_{*}}^{0},
\label{standarTime}
\end{equation}
gives $t=\infty$ for $w<-1$ . For $\omega>-1$ and $1<\alpha\leq2$, eq.~(\ref{singularityProspect1d}) yields
\begin{equation}
t=\frac{\pm1}{3(1+\omega)\sqrt{\pi}}\left(\frac{8\rho_{0}(3\alpha-2)}{3(\alpha+2)}\right)^{\frac{\alpha}{2-3\alpha}}\left(\frac{3\alpha-2}{\alpha}\right)a_{*}^{\frac{3\alpha(1+\omega)}{3\alpha-2}}.
\label{singularityProspect1e}
\end{equation}
This result similarly implies that the time to reach singularity is finite. In fig.~(\ref{sing1}), we present the time to reach singularity with respect to scale factor \footnote{We only present the $\omega=0$ case since similar effects can be seen for radiation$-$dominated and accelerated cases.}. For the larger values of $a$, the time to reach singularity increases while $\alpha$ decreases. For smaller values of $a$, the time to reach singularity increases while $\alpha$ increases. 

\subsection{Generalised second law}
 
GSL states that total entropies of fluid and horizon do not decrease with time. In order to check the validity of GSL, we begin to rearrange eq.~(\ref{dynamicalEquation}) as follows:
 \begin{equation}
 \label{intEq2}
 \dot{\tilde{r_{A}}}=4\pi^{\frac{3\alpha-2}{2\alpha}}\left(\frac{2\alpha}{2+\alpha}\right)(\rho+p)H\tilde{r_{A}}^{\frac{4\alpha-2}{\alpha}}.
 \end{equation}
Combining eqs.~(\ref{TdS}) and (\ref{intEq2}) leads to the following expression
\begin{equation}
\label{GSL}
 T_{h}\dot{S_{h}}=4\pi(\rho+p)H\tilde{r_{A}}^{3}\left[1-2\pi^{\frac{3\alpha-2}{2\alpha}}(\rho+p)\left(\frac{2\alpha}{2+\alpha}\right)\tilde{r_{A}}^{\frac{3\alpha-2}{\alpha}}\right],
\end{equation}
which does not ensure the validity of the second law, .i.e., $S_{h}\geq0$, since $(\rho+p)$ is negative for the accelerated phase. Therefore, we must take into account the entropy of matter fields. To do so, we consider the Gibbs equation which is defined by \cite{Izquierdo2006}
 \begin{equation}
 \label{GE}
 T_{f}dS_{f}=d(\rho V)+pdV=Vd\rho+(\rho+p)dV,
 \end{equation}
 where $T_{f}$ and $S_{f}$ correspond to the temperature and entropy of the matter fields inside the apparent horizon. Moreover, we consider the thermal equilibrium, i.e., $T_{h}=T_{f}$, otherwise the spontaneous energy flow between fluid and horizon would cause the deformation of the FRW geometry \cite{Izquierdo2006}. Furthermore, in order to avoid non-equilibrium thermodynamics, we use the assumption of thermal equilibrium between horizon and fluid. Hence one can obtain the following expression from eqs.  (\ref{continuityEqu}), (\ref{volume}), (\ref{intEq2}) and Gibbs equation
\begin{equation}
\label{GSL2}
T_{h}\dot{S_{f}}=-4\pi\tilde{r_{A}}^{3}(\rho+p)H\left[1-4\pi^{\frac{3\alpha-2}{2\alpha}}(\rho+p)\left(\frac{2\alpha}{2+\alpha}\right)\tilde{r_{A}}^{\frac{3\alpha-2}{\alpha}}\right].
 \end{equation} 
Finally, combining eqs.~(\ref{GSL}) and (\ref{GSL2}), we obtain
  \begin{equation}
 \label{GSLT}
T_{h}(\dot{S_{f}}+\dot{S_{h}})=8\pi^{\frac{5\alpha-2}{2\alpha}}\left(\frac{2\alpha}{2+\alpha}\right)(\rho+p)^{2}H\tilde{r_{A}}^{\frac{6\alpha-2}{\alpha}}.
 \end{equation}
It is clearly obvious that the total entropy change does not decrease with time. Therefore, the GSL always holds for all eras of the universe.

\section{Friedmann equations from entropic gravity}

In this section, we are going to derive the Friedmann equations from Verlinde's entropic gravity. Following the arguments of ref.~\cite{Verlinde2011,Cai2010}, we begin to consider a compact spatial region $\mathcal{V}$ with a holographic screen on the boundary $\mathcal{\partial V}$. We assume that the number of bits on the screen is \cite{Verlinde2011}
\begin{equation}
\label{NARel}
N=A,
\end{equation}
where A is the area of the holographic screen. From entropy-area relation $S=A/4$, N is given by
\begin{equation}
N=4S_{h}.
\label{NSRelation}
\end{equation}
Moreover, we assume that the total energy of the screen obeys the equipartition law. So we have \cite{Verlinde2011}
\begin{equation}
\label{equPart}
E=\frac{1}{2}NT,
\end{equation}
where screen temperature $T$ corresponds to Unruh temperature. It is defined by \cite{Unruh1976,Verlinde2011}
\begin{equation}
\label{unTemp}
T=\frac{a_{r}}{2\pi}=-\frac{\ddot{a}r}{2\pi},
\end{equation}
where $a_{r}$ is the acceleration. It is given by \cite{Cai2010}
\begin{equation}
\label{accl}
a_{r}=-\frac{d^{2}\tilde{r_{A}}}{dt^{2}}=-\ddot{a}r.
\end{equation}
In order to obtain the Friedmann equations, we should consider the active gravitational mass $\mathcal{M}$ instead of the total mass $M$ inside the region $\mathcal{V}$. It is given by \cite{Cai2010}
\begin{equation}
\label{KomarMss}
\mathcal{M}=2\int_{\mathcal{V}}dV\left(T_{\mu\nu}-\frac{1}{2}Tg_{\mu\nu}\right)u^{\mu}u^{\nu}=\frac{4}{3}\pi(\rho+3p)\widetilde{r_{A}}^{3},
\end{equation}
where $T_{\mu\nu}$ is defined as in eq.~(\ref{energyMomentumTensor}). Furthermore, we assume \cite{Verlinde2011}
\begin{equation}
\label{EEqualMass}
E=\mathcal{M}.
\end{equation}

From the fractional entropy (\ref{entropy}), number of bits $N$ (\ref{NSRelation}) is modified as
\begin{equation}
N=4\left(\pi\widetilde{r_{A}}^{2}\right)^{\frac{\alpha+2}{2\alpha}}=4\left(\pi a^{2}r^{2}\right)^{\frac{\alpha+2}{2\alpha}}.
\label{ModifiedBits}
\end{equation}
From eqs.~(\ref{equPart})-(\ref{ModifiedBits}), one can easily derive the second Friedmann equation, i.e., the acceleration equation,
\begin{equation}
\label{FriedmanEntrop1}
\pi^{\frac{2-\alpha}{2\alpha}}\left(\frac{\ddot{a}}{a}\right)\tilde{r_{A}}^{\frac{2-\alpha}{\alpha}}=-\frac{4\pi}{3}(\rho+3p).
\end{equation}
Multiplying $2\Dot{a}a$ on both sides of this equation and using the continuity equation (\ref{continuityEqu}) yields
\begin{equation}
\label{integForfiredmanEntrop2}
\int\frac{d\dot{a}^{2}}{dt}dt=\frac{8\pi}{3}\int\frac{1}{(\pi r^{2})^{\frac{2-\alpha}{2\alpha}}a^{\frac{2-\alpha}{\alpha}}}\frac{d(\rho a^{2})}{dt}dt.
\end{equation}
Choosing again the equation of state as $p=\omega \rho$, we obtain from eq.~(\ref{RhoScaleRel}),
\begin{equation}
 d\left(\rho a^{2}\right)=-\rho_{0}(1+3\omega)a^{-2-3\omega}da.
\label{rhoaRelFinal}
\end{equation}
Using the above expression in  integral (\ref{integForfiredmanEntrop2}), one can obtain
\begin{equation}
\label{FriedmanEntrop2}
\left(\frac{\dot{a}}{a}\right)^{2}+\frac{k}{a^{2}}=\frac{8\pi\rho_{0}}{3}\frac{\alpha(3\omega+1)}{\pi^{\frac{2-\alpha}{2\alpha}}(3\alpha\omega+2)}\frac{a^{-3(\omega+1)}}{\tilde{r_{A}}^{\frac{2-\alpha}{\alpha}}},
\end{equation}
which is a modification of the first Friedmann equation. Using $H=\dot{a}/a$ and eq.~(\ref{apparentHor}), the Friedmann equations are expressed in terms of the Hubble parameters as follows:
 \begin{eqnarray}
 \label{FriedmanEntropH}
 \left(H^{2}+\frac{k}{a^{2}}\right)^{\frac{3\alpha-2}{2\alpha}}=\frac{8\pi\rho}{3}\frac{\alpha(3\omega+1)}{\pi^{\frac{2-\alpha}{2\alpha}}(3\alpha\omega+2)}\nonumber,\\
\pi^{\frac{2-\alpha}{2\alpha}}\left(\dot{H}+H^{2}\right)=-\frac{4\pi}{3}(\rho+3p)\left(H^{2}+\frac{k}{a^{2}}\right)^{\frac{2-\alpha}{2\alpha}},
 \end{eqnarray}
where we used eq.(\ref{RhoScaleRel}) in the first Friedmann equation \footnote{At this point, we give some comments on the first Friedmann equation. In order to evaluate the integral in eq. (\ref{integForfiredmanEntrop2}), we empoly  $p=\omega\rho$. We obtain the first Friedmann equation depending on a specific equation of state.}. Again, in the limit $\alpha\rightarrow2$, the usual Friedmann equations are recovered.

Let us again consider the deceleration parameter $q$ for the flat case. From eqs.~(\ref{decelPar}), (\ref{FriedmanEntrop1}) and (\ref{FriedmanEntrop2}), $q$ is given by
\begin{equation}
q=\frac{2+3\alpha\omega}{2\alpha}.
\label{decelParC2}
\end{equation}
Choosing $\omega=1/3$ and $\omega=0$ gives 
\begin{equation}
q_{rad}=\frac{2+\alpha}{2\alpha},\qquad q_{m}=\frac{1}{\alpha}
\label{decelParRad2}
\end{equation}
for radiation$-$ and matter-dominated eras, respectively. Needless to say, it is clear that the deceleration parameter is reduced to the standard form in the limit $\alpha\rightarrow2$. Similarly, radiation$-$ and matter-dominated eras correspond to deceleration phases since $q>0$. In order to explain the late time acceleration ($q<0$), we again find that $\omega$ must satisfy the condition $ \omega<-\frac{2}{3\alpha}$.

\begin{figure}
\centerline{\includegraphics[width=11cm]{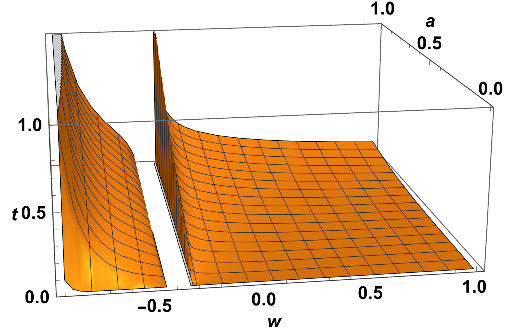}}
\caption{Time to reach  singularity vs scale factor and equation of state parameter. We take $\rho_{0}=1$ and $\alpha=3/2$.}
\label{sing2}
\end{figure}

Lastly, we probe the singular and nonsingular beginning of the universe. To do so, we rearrange the first Friedmann equation in eq.~(\ref{FriedmanEntropH}) as
\begin{equation}
t=\pm\left[\frac{8\pi^{\frac{3\alpha-2}{2\alpha}}(1+3\omega)\alpha\rho_{0}}{3(2+3\alpha\omega)}\right]^{\frac{\alpha}{2-3\alpha}}\int_{a_{*}}^{0}a^{\frac{3(1+\omega)\alpha}{3\alpha-2}-1}da
\label{ReaFried1}
\end{equation}
for the flat case, $k=0$. Calculation of this integral is given by
\begin{equation}
t=\pm\left[\frac{8\pi^{\frac{3\alpha-2}{2\alpha}}(1+3\omega)\alpha\rho_{0}}{3(2+3\alpha\omega)}\right]^{\frac{\alpha}{2-3\alpha}}\frac{3\alpha-2}{3\alpha(1+\omega)}a_{*}^{\frac{3(1+\omega)\alpha}{3\alpha-2}}.
\label{EntropicSingularTime}
\end{equation}
One can see that the result of integral in  eq.~(\ref{ReaFried1}) is finite for $\omega>-1$. Therefore, the result implies that the initial singularity is accessible. However, the prefactor of eq.~(\ref{EntropicSingularTime}) is imaginary in the interval $-\frac{2}{3\alpha}<\omega<-\frac{1}{3}$. So time is also imaginary in this interval. In fig.~(\ref{sing2}), we present the time to reach singularity with respect to the scale factor and the equation$-$of$-$state parameter \footnote{The effects of the fractional parameter resemble fig.~(\ref{sing1}).}.  As can be seen in the figure, time is imaginary for $-4/9<\omega<-1/3$.  Therefore, it is unphysical. In order to avoid imaginary time, we may interpret this result as a constraint on $\omega$. We may exclude the values of $\omega$ between $-2/3\alpha$ and $-1/3$ since time is not defined. We know that standard cosmology does not impose such a constraint on $\omega$. This unexpected constraint arises with fractional modification. This interval may imply that the universe will not arise or cosmic beginning will not occur.

Finally, we give some comments on the consistency of the two methods since one may expect the two methods to have the same results. In contrast to the previous section, we derived the Friedmann equations depending on a specific equation of state in this section. Hence, eqs. (\ref{Fried2}) and (\ref{FriedmanEntropH}) are not exactly the same. Nevertheless, if we absorb the extra terms in $\rho$, the first Friedmann equations can be regarded as the same.

\section{Conclusions}

 In this paper, using fractional entropy \cite{Jalalzadeh2021}, we obtained fractional modified Friedmann equations from two different approaches. First, we derived the Friedmann equations from the first law of thermodynamics at apparent horizon \cite{Akbar2007}. We checked the validity of GSL and showed that it is valid for all eras. Then, we investigated the effects of fractional parameters on the deceleration parameter and time to reach the initial singularity. We found that $q$ is still positive for radiation and matter-dominated eras. Therefore, matter$-$ and radiation-dominated eras correspond to the deceleration phase. The results indicate that fractional effects cannot provide an alternative to dark energy. So we still need dark energy to explain the late time acceleration. Moreover, $\omega$ must obey the condition $\omega<-\frac{2}{3\alpha}$ for $q<0$ at late time acceleration. Based on the method in ref.~\cite{Awad2013,Hendi2017}, we also calculated the time to reach the initial singularity from a finite initial density $\rho_{*}$ to the Big Bang density $\rho\rightarrow\infty$. Moreover, we repeat our investigation for the scale factor, namely we calculate the time to reach the initial singularity from a nonzero scale factor $a_{*}$ to $a\rightarrow0$. Our analysis reveals that initial singularity is accessible since time is finite.  

Next, we derived modified Friedmann equations from Verlinde's entropic gravity approach \cite{Verlinde2011,Cai2010}. Again, we handle the deceleration parameter and time to reach the initial singularity. Just like the results in the previous section, we found similar results for the deceleration parameters. Finally, we investigated the time to reach the initial singularity in the entropic gravity case. The result indicates that the singularity is accessible. We found that time is imaginary, namely, unphysical for $-\frac{2}{3\alpha}<\omega<-\frac{1}{3}$. Therefore, we may interpret the result as a constraint on $\omega$. 

For the sake of completeness, we also mention the papers in refs. \cite{Genarro2022,Genarro2022b} where the authors explored cosmological
consequences and modified gravity for Barrow entropy.

\section*{Acknowledgments}
The authors thank the anonymous referees for their helpful and constructive comments. This work was supported by Istanbul University Post-Doctoral Research Project: MAB-2021-38032.

Data availability statement: No new data were created or analysed in this study.

\end{document}